\documentclass[twocolumn,pra,superscriptaddress,longbibliography]{revtex4-2}

\usepackage{amsmath,amsfonts,amssymb}
\usepackage{graphicx,color}
\usepackage{comment}
\usepackage[breaklinks=true]{hyperref}
\usepackage{breakcites}
\usepackage{epstopdf}
\usepackage{float}
\usepackage{multirow}
\usepackage{hhline}
\usepackage[normalem]{ulem}
\usepackage{mathtools}
\usepackage[caption=false]{subfig}
\usepackage{soul}
\hypersetup{hidelinks,colorlinks=true,allcolors=BLUE}
\usepackage{physics}
\usepackage{nicefrac}
\usepackage{bbold}

\newcommand{\ham}{\hat{\mathcal{H}}}
\newcommand{\p}{\hat{p}}
\newcommand{\x}{\hat{x}}
\newcommand{\n}{\hat{n}}
\newcommand{\kl}{k_L}
\renewcommand{\wr}{\omega_R}
\newcommand{\phit}{\varphi(t)}
\renewcommand{\b}{\hat{b}}

\newcommand{\infid}{\mathcal{I}}
\newcommand{\loss}{\mathcal{L}}

\makeatletter

\@ifundefined{textcolor}{}
{%
 \definecolor{BLACK}{gray}{0}
 \definecolor{WHITE}{gray}{1}
 \definecolor{RED}{rgb}{1,0,0}
 \definecolor{GREEN}{rgb}{0,1,0}
 \definecolor{BLUE}{rgb}{0,0,1}
 \definecolor{CYAN}{cmyk}{1,0,0,0}
 \definecolor{MAGENTA}{cmyk}{0,1,0,0}
 \definecolor{YELLOW}{cmyk}{0,0,1,0}
}

\makeatother

\newcommand{\Infleqtion}{Infleqtion, Inc., Chicago, Illinois 60604, USA}
\newcommand{\UChicago}{Department of Computer Science, University of Chicago, Chicago, Illinois 60637, USA}
\newcommand{\Infleqtionboulder}{
Infleqtion, Inc., 3030 Sterling Circle, Boulder, Colorado 80301, USA}

\begin{document}

\title{Programming an Optical Lattice Interferometer}

\author{Lennart Maximilian Seifert}
\email[Electronic address: ]{lmseifert@uchicago.edu }
\affiliation{\UChicago}

\author{Victor E.~Colussi}
\affiliation{\Infleqtionboulder}

\author{Michael A.~Perlin}
\affiliation{\Infleqtion}

\author{Pranav Gokhale}
\affiliation{\Infleqtion}

\author{Frederic T.~Chong}
\affiliation{\UChicago}
\affiliation{\Infleqtion}

\date{\today}

\begin{abstract}
Programming a quantum device describes the usage of
quantum logic gates, agnostic of hardware specifics, to perform a sequence of operations with (typically) a computing or sensing task in mind.
Such programs have been executed on digital quantum computers, which despite their noisy character, have shown the ability to optimize metrological functions, for example in the generation of spin squeezing and 
optimization
of quantum Fisher information for signals manifesting as spin rotations in a quantum register.
However, the qubits of these programmable quantum sensors are tightly spatially confined and therefore suboptimal for enclosing the kinds of large spacetime areas required for performing inertial sensing.
In this work, we derive a set of quantum logic gates for a cold atom optical lattice interferometer that manipulates the momentum of atoms.
Here, the operations are framed in terms of single qubit operations and mappings between qubit subspaces with internal levels given by the Bloch (crystal) eigenstates of the lattice.
We describe how the quantum optimal control method of direct collocation is well suited for obtaining the modulation waveforms of the lattice which achieve these operations.
\end{abstract}

\maketitle

\section{Introduction}

Advancements in laser and vacuum technologies over the last several decades have enabled the trapping, cooling, and precise manipulation of atoms in the lab.  These setups typically operate at sub-microkelvin temperatures where quantum properties such as coherence, superposition, and entanglement are manifest and can be manipulated and leveraged.  Consequently, ultracold atoms are now at the forefront in the development of quantum technologies in the computing, simulation, and sensing sectors (c.f.~\cite{degenQuantumSensing2017b, grossQuantumSimulationsUltracold2017a, laddQuantumComputers2010a} and refs therein).

In the realm of digital quantum computing, the language of quantum logic gates has been developed to describe universal sets of operations that may be performed on a quantum computer (regardless of the platform) in order to implement quantum algorithms \cite{nielsenQuantumComputationQuantum2010b}.  Such algorithms consist of state preparation, single-qubit rotations, entangling blocks, mid-circuit measurements, readout, etc., based on single, two, and multi-qubit operations.  Although this paradigm is intended for performing digital quantum computing, the 
high degree of control over quantum systems
has opened up analog applications including quantum simulation of many-body Hamiltonians (particularly of lattice systems) \cite{georgescuQuantumSimulation2014b, daleyPracticalQuantumAdvantage2022a} and on-demand generation of quantum resources for quantum sensing \cite{marciniakOptimalMetrologyProgrammable2022a, kaubrueggerVariationalSpinSqueezingAlgorithms2019b}.  In the latter case, quantum computers have been explored as programmable quantum sensors, mainly exploring the preparation of states according to metrological cost functions.  

These developments motivate the paradigm of ``programmable'' quantum sensing, which describes using quantum logic gates, in a hardware agnostic manner, to perform sensing tasks.  Much attention has been paid to using NISQ machines as quantum sensors of spin rotations due to, for example, stray electromagnetic fields \cite{marciniakOptimalMetrologyProgrammable2022a, kaubrueggerVariationalSpinSqueezingAlgorithms2019b, castroVariationalQuantumState2024a, kaubrueggerOptimalVariationalMultiparameter2023a}.  However, due to the tight confinement of qubits in quantum computing platforms, these machines are ill-suited for measuring inertial and gravitational forces, limiting their application as programmable inertial sensors.  Conventional cold-atom inertial sensing platforms are instead able to generate large momentum transfers and subsequently split clouds of atoms to enclose large spacetime areas, thereby yielding high sensitivities \cite{kasevichAtomicInterferometryUsing1991a, mullerAtomInterferometryTestsIsotropy2008a, kovachyQuantumSuperpositionHalfmetre2015b}.  The set of possible operations in such systems is however constrained due to the lack of reconfigurability of the platforms and the restriction to limited spatial regions of a vacuum chamber where the atoms can be addressed, which severely limits their programmability.

Recent experiments have however demonstrated that cold-atom inertial sensing can be performed using optical lattices, where the atoms remained always confined in a standing wave of light, and the basic steps of an interferometer--splitting, mirroring, and recombining--can be achieved by time-dependent modulation of the phase of the light forming the lattice \cite{weidnerExperimentalDemonstrationShakenLattice2018a, ledesmaDemonstrationProgrammableOptical2024, ledesmaVectorAtomAccelerometry2024a}.  Because the atoms remain always confined (albeit delocalized), the modulation waveforms for efficiently generating the traditional components required for interferometry are non-trivial.  Previous studies have outlined how machine-learning methods, including reinforcement learning and (indirect) quantum optimal control, can be used to perform state mappings and design the traditional components (split, mirror, recombine) as well as find modulation sequences spanning the entire operation time of the sensor which maximize the overall sensitivity \cite{ledesmaDemonstrationProgrammableOptical2024, pottingMomentumstateEngineeringControl2001c, dupontQuantumStateControl2021b, weidnerAtomInterferometryUsing2017b, weidnerSimplifiedLandscapesOptimization2018b, chihReinforcementlearningbasedMatterwaveInterferometer2021c, shaoApplicationQuantumOptimal2023b, nicotraModelingControlUltracold2023b, alamRobustQuantumSensing2024a, chihReinforcementLearningRotation2024, chihMachineLearningBasedDesignQuantumg}.  
While much attention has been paid to these traditional component and end-to-end design problems, the reconfigurability of the optical lattice interferometer can also be used to perform a wider range of operations, making it a natural candidate for an inertial quantum sensing platform which can execute a broad array of user-defined programs.

In this paper, we use a direct trajectory-based quantum optimal control method to design control functions implementing various quantum logic gates for the optical lattice interferometer system.  The direct trajectory optimization method, which has thus far been applied towards gate design for quantum computing \cite{trowbridgeDirectCollocationQuantum2023b}, has distinct advantages including the ability to enforce constraints on the evolution of the quantum state and design cost functions that require information from intermediate time steps.  We find that this method is able to produce high-fidelity gates, for instance for large momentum two-level (qubit) subspaces. 

The paper is structured as follows.
We introduce the optical lattice atomic interferometer and the direct trajectory optimization method in Sec.~\ref{sec:background}.
In Sec.~\ref{sec:opt}, we derive and analyze a basic set of logic gates operating within and between qubit subspaces, given by the Bloch bands of the lattice, and give the corresponding modulation waveforms that achieve the desired quantum operation to high fidelity.
Finally, we summarize our work in Sec.~\ref{sec:conclusion}, and provide an outlook on the constructions of gate circuits for inertial sensing.

\section{Background}\label{sec:background}

\begin{figure}[t!]
    \centering
    \includegraphics[width=0.9\linewidth]{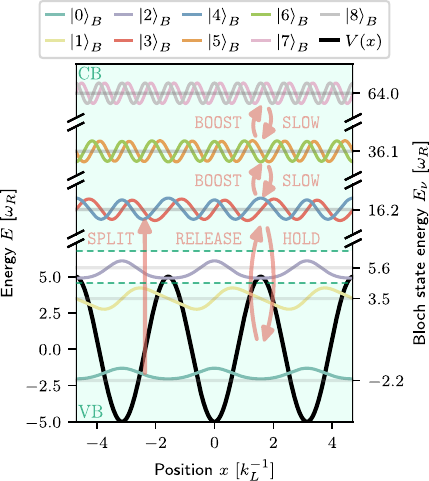}
    \caption{Band structure of the optical lattice Hamiltonian (Eq.~\eqref{eq:ham}) for $\varphi \equiv 0$. The potential $V(x)$ for the specific choice $V_0 = 10 \, \wr$ is shown in black, while modulating $\varphi$ corresponds to shifting the potential to the left or right. The Bloch states $\psi_\nu(x) = \ip{x}{\nu}_B$ are shown at their corresponding energy eigenvalue, clearly defining a valence and a conduction band subspaces, with latter characterized by nearly degenerate pairs of opposite-parity states. A subset of the operations considered in this work are indicated by red arrows (see also Sec. \ref{sec:opt} and Fig. \ref{fig:bloch_spheres}).}
    \label{fig:spectrum}
\end{figure}

In this section, we describe both the cold-atom optical lattice platform (see Sec.~\ref{sec:sl}), as well as quantum optimal control method used to design quantum operations (see Sec.~\ref{sec:qoc}) in this system.  The description of cold atoms in an optical lattice is kept at a basic level, and we refer interested readers to Ref.~\cite{ledesmaDemonstrationProgrammableOptical2024,ledesmaProgrammableBlochbandAtom2024a} for further details of this system, including the implementation of state preparation, control, and measurement in the lab.

\subsection{Optical lattice interferometry}
\label{sec:sl}
We consider a cloud of cold, spinless bosonic atoms in a one-dimensional optical lattice that is generated by two counter-propagating laser beams, as realized in Refs.~\cite{weidnerExperimentalDemonstrationShakenLattice2018a,ledesmaDemonstrationProgrammableOptical2024}.  Furthermore, we consider density regimes where interaction effects between atoms can be neglected, reducing the description to the single-atom level.  At sufficiently low temperatures, the cloud becomes Bose-condensed such that the wave function of the entire system of $N$ atoms can be represented as a product $\Psi(x_1,\dots,x_N,t) = \prod_{i=1}^N\psi(x_i,t)$ of the single-particle wave function $\psi(x,t)$.  The wave function $\psi(x,t)$ evolves according to the Schr\"odinger equation with time-dependent Hamiltonian
\begin{align}
    \label{eq:ham}
    \ham(t) = \frac{\p^2}{2m} + \hat{V}(\x, t),
\end{align}
where $\hat p$ and $\hat x$ are, respectively, the momentum and position operators for an atom with mass $m$, and
\begin{align}
    \hat{V}(\x, t) =  - \frac{V_0}{2} \cos(2 \kl \x + \phit),
\end{align}
is the potential generated by the optical lattice at time $t$.  The depth $V_0$ and wavenumber $\kl$ of the lattice are held fixed while its phase $\phit$ is modulated according to a time-dependent control function whose waveform is determined using quantum optimal control as described in Sec.~\ref{sec:qoc}.  Previous works have developed control functions implementing split, mirror, and recombine components of an interferometer \cite{ledesmaDemonstrationProgrammableOptical2024, chihReinforcementlearningbasedMatterwaveInterferometer2021c, shaoApplicationQuantumOptimal2023b, nicotraModelingControlUltracold2023b} as well as end-to-end protocols maximizing the overall sensitivity of the device \cite{alamRobustQuantumSensing2024a, chihReinforcementLearningRotation2024}.

The lattice potential can be written in the momentum basis as
\begin{align}
    \hat{V}(t) = - \frac{V_0}{4} \qty(e^{i \phit} \b_- + e^{-i \phit} \b_+),
\end{align}
where the operator $\b_\pm$ is defined by $\b_\pm|p\rangle  = |p \pm 2\kl\rangle$ (as usual, $\hbar=1$), representing a two-photon process in which an atom absorbs and emits a photon between counterpropagating laser beams, changing its momentum by $\pm 2\kl$.  In the case of a stationary lattice ($\varphi \equiv 0$), the eigenstates of the lattice Hamiltonian are given by the Bloch states $\ket{\nu,q}$ with eigenergies $E_\nu(q)$ parametrized by the band index $\nu$ and quasimomentum $q$ (c.f.~\cite{kittelIntroductionSolidState2005, pitaevskiiBoseeinsteinCondensationSuperfluidity2016b}).  The Bloch states are of alternating parity with spatial dependence and eigenenergies shown in Fig.~\ref{fig:spectrum}.  In particular, the quasimomentum is left unchanged by the action of $\hat{V}$ regardless of the form of the control function $\varphi(t)$.  The application of a linear force with Hamiltonian contribution $F\hat{x}$ instead leads to a time-dependent $q(t)$.

 In general, an inertial force leads to a final state with different expansion coefficients and quasimomentum than in an (ideal) unaccelerated frame.  Experimentally, the cloud is split, mirrored, and then recombined using modulated waveforms of the control function, with each component separated by a period of free evolution.  Measurement is performed following the termination of the control function, and the atomic cloud is allowed to ballistically expand for a fixed amount of time.  This leads asymptotically to a separation of the wave function into the basis of discrete momentum states of the lattice $\{|2n\kl + q\rangle\}_{n\in \mathbb{Z}} $ where $\mathbb{Z}$ is the set of integers \cite{ledesmaDemonstrationProgrammableOptical2024}.  Depending on the value of $F$, different populations of the discrete momentum states are observed, producing a multi-fringed interference pattern that reflects differential phase accumulation between arms of the interferometer. This is due to the possibility of occupying an array of discrete momentum states at integer multiples of the lattice wavenumber. In this work, we find control functions for the case $F=0$, which can be straightforwardly generalized to $F\neq 0$ via the application of an offsetting bias field or through re-optimization. 

It is convenient to work with the natural units of the lattice, where spatial dimensions are defined via the wave number $\kl$ and energy scales are determined by the photon recoil frequency $\wr = \kl^2/2m$. In these units, the momentum-space representation of the Hamiltonian reads ($q=0$):
\begin{align}
    \label{eq:sl_ham}
    \frac{\ham(t)}{\wr} = 4\n^2 - \frac{V_0}{4} \qty(e^{i \phit} \b_- + e^{-i \phit} \b_+).
\end{align}
Furthermore, because $q$ is a constant of motion, we use the notations $|2n \kl\rangle_p$ for the discrete momentum states (with $\hat{n}^2 \ket{2n\kl}_p = n^2 \ket{2n\kl}_p$) and $|\nu\rangle_{B} \equiv |\nu,q=0\rangle$ for the Bloch states. Note that in this representation the operators $\hat{b}_\pm$ effectively perform the mapping $n \rightarrow n \pm 1$.

For concreteness, we fix $V_0 = 10 \, \wr$, which is a representative value of lattice depths used experimentally \cite{ledesmaDemonstrationProgrammableOptical2024}. We note that the results in this work are specific to this choice of the lattice depth but the presented framework can be applied to any other setting in a straightforward fashion. As shown in Fig.~\ref{fig:spectrum}, the Bloch states $\ket{0}_B$ and $\ket{1}_B$ have harmonic oscillator-like wave functions that are well localized in the lattice potential and thus form the {\it valence band}. The higher-energy {\it conduction band} ($\nu \geq 3$) features quadratic dispersions with opposite parity wave functions in the (approximate) form of plane waves.  As the naming convention suggests, atoms in the conduction bands translate under the application of a weak external force, while atoms in the valence bands undergo (Bloch) oscillations \cite{pitaevskiiBoseeinsteinCondensationSuperfluidity2016b, bendahanBlochOscillationsAtoms1996a}. The state $\ket{2}_B$ can be viewed as an intermediate state that with properties similar to both bands.

Formally, the fixed-$q$ Bloch bands illustrated in Fig.~\ref{fig:spectrum} form a qudit, with conduction and valence bands being infinite and finite-dimensional subspaces, respectively.  In practice, atoms are prepared in the ground Bloch state $|\nu = 0\rangle_B$, and then the dynamics of the control function drive metrologically useful transitions between various bands prior to measurement.  Of particular importance are the (nearly) degenerate pairs of opposite-parity states in the conduction band.  The sum and difference of these states correspond (approximately) to individual momentum eigenstates, i.e. $(|\nu\rangle_B \pm |\nu-1\rangle_B)/\sqrt{2} \approx \ket{\pm \nu \kl}_p$.  Given our specific choice of the lattice depth $V_0$, we identify these pairs for even $\nu \geq 4$ and as can be seen in Fig.~\ref{fig:spectrum}. Here the parity of the state $\ket{\nu}_B$ is equivalent to the parity of the integer $\nu$. 

A qubit subspace can be constructed from any pair of Bloch states, however we will find it useful to take neighboring pairs $\qty(|\nu\rangle_B, |\nu -1\rangle_B)$ that are defined by projectors
\begin{align}
    \hat{\Pi}_\nu = |\nu\rangle_B\langle\nu|_B + |\nu-1\rangle_B\langle \nu-1 |_B.
\end{align}
For a given subspace $\hat{\Pi}_\nu$ we denote the logical qubit isometry in these subspaces as
\begin{align}
    \hat{P}_\nu = \ket{0}^\nu_L \! \bra{\nu}_B + \ket{1}^\nu_L \! \bra{\nu-1}_B,  
\end{align}
where we choose the convention that the even-parity state is assigned to the north pole of the logical Bloch sphere, associated with $\ket{0}_L$.
For later convenience, the qubit subspace of the ground and first excited Bloch states are labeled following a different convention $\hat{\Pi}_1 = |0\rangle_B\langle 0|_B + |1\rangle_B\langle 1|_B$ (isometry $\hat{P}_1 = \ket{0}^1_L \! \bra{0}_B + \ket{1}^1_L \! \bra{1}_B$).

In Sec.~\ref{sec:opt}, we describe gatesets within and between these qubit subspaces. We now introduce a quantum optimal control method for designing control functions that produce gates.

\subsection{Quantum optimal control}
\label{sec:qoc}
Given a closed quantum system with a Hamiltonian of the form
\begin{align}
    \label{eq:qoc_ham}
    \ham(t) = \ham_0 + \sum_{j=1}^{m} \gamma_j(t) \ham_j,
\end{align}
where $\ham_0$ denotes the drift Hamiltonian and $\gamma_j(t)$ represent the control functions coupling to drive Hamiltonians $\ham_j$, quantum optimal control is a numerical framework to find control functions that make the system undergo a desired transformation. This can be a state transfer or unitary transformation for instance. A number of toolboxes have been developed for this task \cite{trowbridgeDirectCollocationQuantum2023b, peterssonOptimalControlClosed2022, goerzKrotovPythonImplementation2019b, canevaChoppedRandombasisQuantum2011b, khanejaOptimalControlCoupled2005b}. Generally they find a solution by minimizing a loss function $\loss$, which is typically primarily given by the operation's infidelity $\infid$. For a state transfer we have
\begin{align}
    \label{eq:infid_state}
    \infid = 1 -  \abs{\ip{\phi}{\psi(T)}}^2,
\end{align}
where $\ket{\psi(T)}$ is the quantum state at the final time $T$ and $\ket{\phi}$ is the goal state, while for a unitary transformation the common choice for the infidelity is related to the Hilbert-Schmidt inner product:
\begin{align}
    \label{eq:infid_unitary}
    \infid = 1 - \frac{1}{d^2}\abs{\Tr[\hat{W}^\dagger \hat{U}(T)]}^2.
\end{align}
Here $d$ denotes the Hilbert space dimension, $\hat{W}$ the target unitary and $\hat{U}(T)$ is the unitary induced by the control functions.

Past works have discussed gradient-based tools \cite{peterssonOptimalControlClosed2022, goerzKrotovPythonImplementation2019b, canevaChoppedRandombasisQuantum2011b, khanejaOptimalControlCoupled2005b} as well as reinforcement learning approaches (in particular in the optical lattice context) \cite{ledesmaDemonstrationProgrammableOptical2024,chihReinforcementlearningbasedMatterwaveInterferometer2021c} to solve the quantum control problems. In most cases these works proposed so-called {\it indirect} methods, where the Schrödinger equation is solved numerically for each choice of control parameters, and the control parameters are adjusted iteratively to decrease the loss function. While this is a memory-efficient approach to the optimal control problem, it is restrictive if one wishes to also enforce constraints on the quantum state evolution or design cost functions that require information from intermediate time steps. Furthermore, reinforcement learning methods are usually limited by their discrete, pre-defined action space, which typically leads to lower fidelities and requires further knowledge to determine the set of actions.


In this work we look at a different approach known as {\it direct trajectory optimization}, which treats the quantum states at all times as free parameters, and treats equations of motion as constraints on these parameters.
Minimization of the loss function can then be performed via interior-point optimization methods.
While this approach is more memory-intensive, as it must store the entire space-time history of a quantum state (or unitary) trajectory, it allows for a high degree of flexibility for enforcing, for example, constraints on the state trajectory or bandwidth constraints on the control functions. Here we use the \texttt{Piccolo.jl} framework \cite{trowbridgeDirectCollocationQuantum2023b}.

We denote the ``state'' at the discrete time $t_l$ with $x_l$, which corresponds to either the quantum state $\ket{\psi_l}$ (for state transfer) or the unitary $\hat{U}_l$ (for gate design). In this setup, $x_l$ become additional optimization variables in an extended nonlinear constrained optimization problem:
\begin{gather}
    x_l, \gamma_{jl} = \arg\min_{x_l, \gamma_{jl}} \loss(x_l, \gamma_{jl}) \\
    \begin{aligned}
        \label{eq:SE_con}
        \text{subject to:} \quad &x_{l+1} - \exp(- i \ham(\gamma_l) \Delta t) x_l = 0, \\
            & x_0 = x(0)
    \end{aligned}
\end{gather}
where $\Delta t$ is the discrete time step. In the simplest case, the loss is given by the infidelity, $\loss(x_l, \gamma_{jl}) = \infid(x_{N_t})$. The advantage of this approach is that the cost function explicitly depends on a subset of optimization variables, which allows for trivial gradient computation of $\loss$ in these variables. The full gradient step for all variables $x_l$ and $\gamma_{jl}$ is then furthermore determined by constraints like those in Eq.~\eqref{eq:SE_con}. Moreover, this approach gives us direct access to the quantum state or unitary at intermediate times of the evolution, which we will exploit when designing control function waveforms. We refer the interested reader to Ref. \cite{trowbridgeDirectCollocationQuantum2023b} for further details.

In its current version \texttt{Piccolo.jl} supports Hamiltonians of the form shown in Eq.~\eqref{eq:qoc_ham}, which is not directly compatible with the Hamiltonian in Eq.~\eqref{eq:sl_ham} as the desired control function $\varphi(t)$ enters nonlinearly. Therefore we introduce auxiliary quadrature controls
\begin{align}
    \label{eq:IQ}
    I(t) = \cos(\varphi(t)), \qquad Q(t) = \sin(\varphi(t)),
\end{align}
so that we obtain
\begin{align}
    \ham(t) = 4 \n^2 -\frac{V_0}{4} \qty(I(t) \qty(\b_- + \b_+) + i Q(t) \qty(\b_- - \b_+)).
\end{align}
After discretization, the quantum optimal control problem then takes the form 
\begin{align}
    \label{eq:opt_prob}
    \hat{x}_l, \hat{I}_{l}, \hat{Q}_{l}, \hat{\varphi}_{l} = \arg\min_{x_l, I_{l}, Q_{l}, \varphi_{l}} \loss(x_l, \varphi_l)
\end{align}
subject to the constraints in Eqs.~\eqref{eq:SE_con} and \eqref{eq:IQ} as well as
\begin{align}
    \label{eq:phi_con}
    \varphi \in [-\pi, \pi], \quad \varphi_0 = \varphi(0) = \varphi_{N_t} = \varphi(T) = 0,
\end{align}
which constrains the range of the control function and resets the position of the lattice at the final time $t=T$.  

In this current formulation there is nothing constraining the shape of the control function $\varphi(t)$. However, for practical reasons of implementation on hardware, we do not want it to be discontinuous or oscillating too strongly. To this end we augment the cost function by a term
\begin{align}
    \loss_\omega[\varphi] &= \int_0^T \abs{\varphi(t) - \int_0^T K(t-s) \varphi(s) \dd{s}}^2 \dd{t} \notag \\
    \overset{\text{discretize}}{\rightarrow} \loss_\omega(\varphi_l) &= \Delta t \sum_{l=1}^{N_t} \abs{\varphi_l - \Delta  t \sum_{m=1}^{N_t} K_{lm} \varphi_m}^2,
\end{align}
that penalizes the Fourier spectrum $\varphi(\omega)$ beyond a cutoff $\omega_c$, which we choose to be $\omega_c = 70 \, \wr$. $K_{lm} = K(t_l - t_m)$ is the convolution kernel and here we use a $\mathrm{sinc}$ kernel:  
\begin{equation}
    K(t) = \frac{\omega_c}{\pi} \operatorname{sinc}(\omega_c t).
\end{equation}

The final form of the optimization problem is now given by Eq.~\eqref{eq:opt_prob} with constraints given by Eqs.~\eqref{eq:SE_con}, \eqref{eq:IQ}, \eqref{eq:phi_con} and $\loss(x_l, \varphi_l) = \infid(x_{N_t}) + 100 \, \loss_\omega(\varphi_l)$. This is used to find the results in Sec. \ref{sec:opt}, and specific cases where more constraints are added to improve the convergence of the numerical optimization will be addressed as encountered.
\section{Gate Design}
\label{sec:opt}

\begin{figure*}[htbp]
    \centering
    \includegraphics[width=0.9\linewidth]{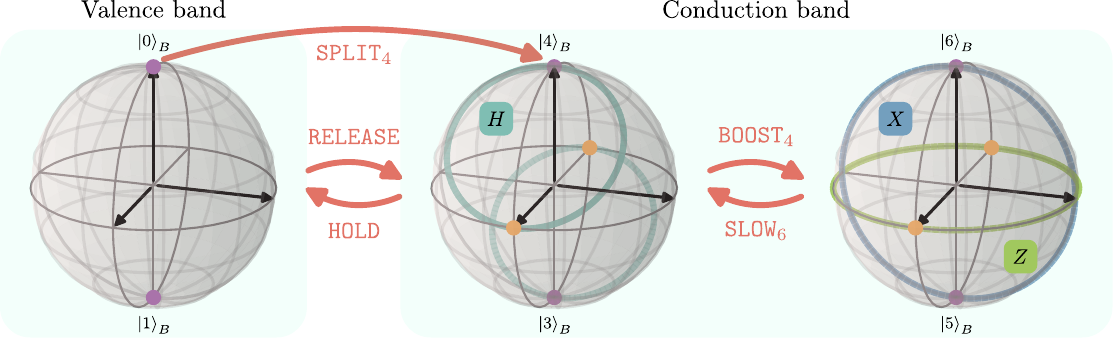}
    \caption{Visualization of a variety of operations that can be realized in the optical lattice interferometer. Neighboring pairs of Bloch states are identified as qubit subspaces and represented by separate Bloch spheres. Upon initialization the system is in $\ket{0}_B$ in the valence band ($\hat\Pi_1$ qubit). \texttt{SPLIT}$_\nu$ maps to a Bloch state $\ket{\nu}_B$ in the conduction band, here shown for $\nu = 4$. Unitary \texttt{BOOST}$_\nu$/\texttt{SLOW}$_\nu$ operations map between adjacent conduction band qubits, for instance \texttt{BOOST}$_4 : \hat\Pi_4 \rightarrow \hat\Pi_6$ and \texttt{SLOW}$_6 : \hat\Pi_6 \rightarrow \hat\Pi_4$. The mapping $\hat\Pi_4 \rightarrow \hat\Pi_1$ corresponds to bringing the atoms to rest (\texttt{HOLD} operation, inverse is \texttt{RELEASE}).  Operations within the group $SU(2)$ on the conduction band qubits are also shown with interpretation in terms of atom-cloud trajectories given in Sec.~\ref{sec:qubitgate}. The yellow circles in the conduction band Bloch spheres correspond to approximate momentum eigenstates $\ket{\pm \nu \kl}_p$.}
    \label{fig:bloch_spheres}
\end{figure*}

In this section, we analyze the results for control functions producing the types of operations shown in Fig.~\ref{fig:bloch_spheres}. Here, qubit subspaces of interest have been represented by individual Bloch spheres as well as mapping operations between these subspaces.  The mapping operations are also visualized in Fig.~\ref{fig:spectrum} for reference.  We comment on the relevance of each operation for inertial sensing. 

\subsection{\texttt{SPLIT}$_\nu$, \texttt{RECOMBINE}$_\nu$}
\label{sec:split}

The Mach-Zehnder interferometer is a paradigmatic sensing scheme comprised of \texttt{SPLIT}$_\nu$, \texttt{MIRROR}$_\nu$ and \texttt{RECOMBINE}$_\nu$ operations.  Given the central importance of this protocol we first analyze the results for the control function that produces the \texttt{SPLIT}$_\nu$ action. This operation maps the ground state $|0\rangle_B$, which corresponds to the initial state, to a Bloch state $\ket{\nu}_B$ in the conduction band. We note that this operation has been analyzed in several related works specifically for the choice $\nu = 3$, each finding different control functions \cite{ledesmaDemonstrationProgrammableOptical2024, chihReinforcementlearningbasedMatterwaveInterferometer2021c, shaoApplicationQuantumOptimal2023b}. As noted previously \cite{chihReinforcementlearningbasedMatterwaveInterferometer2021c}, the inverse \texttt{RECOMBINE}$_\nu$ operation can be achieved through the time reversal of \texttt{SPLIT}$_\nu$.

\begin{figure}
    \centering
    \includegraphics[width=0.9\linewidth]{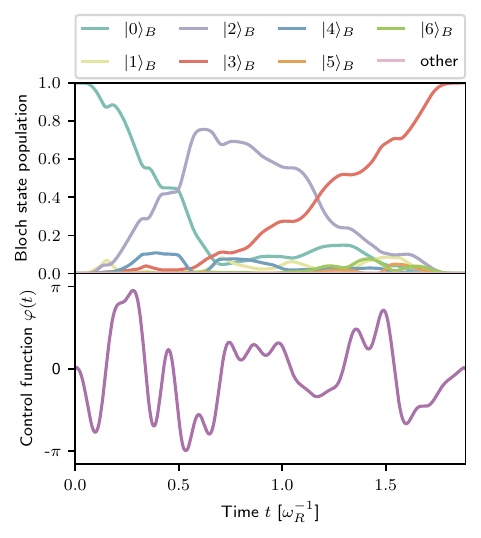} 
    \caption{\texttt{SPLIT}$_3 : \ket{0}_B \rightarrow \ket{3}_B$  operation, $T = 1.88 \, \wr^{-1}$, $\infid = 3.27 \times 10^{-5}$. The top panel shows the state evolution in the Bloch basis while the lower panel shows the corresponding control function.}
    \label{fig:split}
\end{figure}

We briefly present our result for this well-studied task. In order to optimize the desired state transfer, we consider the optimization problem in Eq.~\eqref{eq:opt_prob} with infidelity function Eq.~\eqref{eq:infid_state} and $\ket{\psi(0)} = \ket{0}_B$, $\ket{\phi} = \ket{3}_B$. Figure~\ref{fig:split} shows the obtained control function together with the Bloch state evolution. The even-parity $\ket{2}_B$ state is heavily populated at intermediate times as population is transported between valence and conduction bands.

\subsection{Qubit gates}\label{sec:qubitgate}
\begin{figure*}[htbp]
    \centering
    \includegraphics[width=\linewidth]{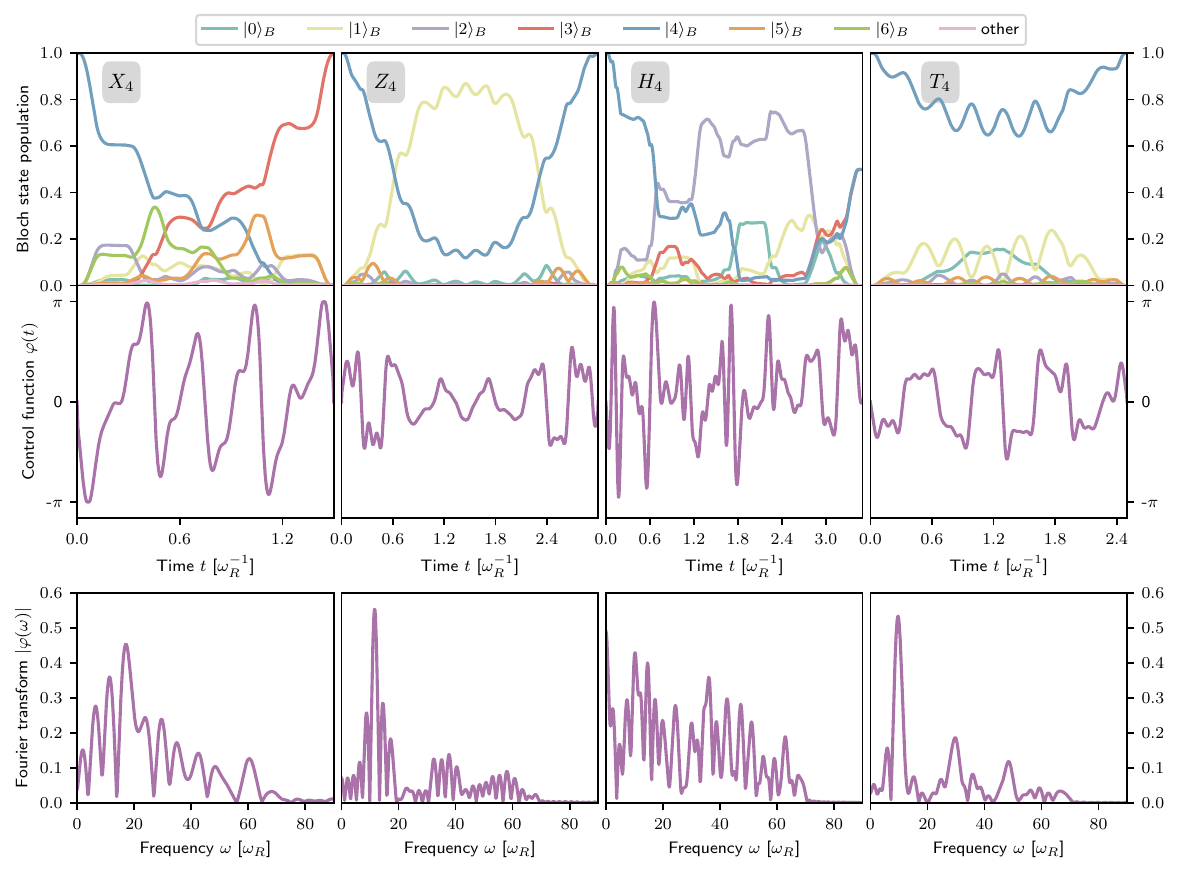}
    \caption{Realization of logical qubit gates on the $\hat{\Pi}_4$ qubit. Each column shows the state evolution of one basis state, the control function as well as its Fourier transform. The desired frequency cutoff at $\omega = 70 \wr$ is achieved through an additional term in the objective and enables smooth waveforms. The gate properties $\qty(\text{duration }T, \text{infidelity } \infid)$ are $\qty(1.50 \, \wr^{-1}, 4.05 \times 10^{-3})$ for $\hat{X}_4$, $\qty(3.00 \, \wr^{-1}, 2.13 \times 10^{-3})$ for $\hat{Z}_4$, $\qty(3.50 \, \wr^{-1}, 9.39 \times 10^{-5})$ for $\hat{H}_4$ and $\qty(2.50 \, \wr^{-1}, 3.61 \times 10^{-4})$ for $\hat{T}_4$.}
    \label{fig:xzht}
\end{figure*}

Within a qubit subpace $\hat{\Pi}_\nu$, arbitrary unitary transformations can be performed within the group $SU(2)$.  In Fig.~\ref{fig:xzht}, we provide results for the control functions achieving the gates
\begin{equation}\label{eq:qubitgates}
    \begin{gathered}
        \hat{X} = \mqty(0&1\\1&0) \qc \hat{Z} = \mqty(1&0\\0&-1), \\
        \hat{H} = \frac{1}{\sqrt{2}} \mqty(1&1\\1&-1) \qc \hat{T} = \mqty(1&0\\0&e^{i \pi/4}).
    \end{gathered}
\end{equation}
$\hat{X}$ and $\hat{Z}$ are the Pauli gates, $\hat{H}$ is the Hadamard gate and $\hat{T} = \hat{Z}^{1/4}$.  It is well known that any element of $SU(2)$ can be approximated by a product of $\hat{H}$ and $\hat{T}$ gates \cite{nielsenQuantumComputationQuantum2010b}. We equip a gate label with the index $\nu$ to denote the qubit subspace it acts on, for instance $\hat{X}_\nu$ acts on qubit subspace $\hat\Pi_\nu$.

Although this gateset is ubiquitous in circuit models of quantum computing (c.f.~\cite{nielsenQuantumComputationQuantum2010b}), we provide their interpretation and utility for operations within the the conduction band in the context of inertial sensing. First, the $\hat{Z}_\nu$ corresponds to the \texttt{MIRROR}$_\nu$ component in the Mach-Zehnder interferometer, which inverts the momenta of the interferometer arms. This can be seen by looking at the $x$-axis eigenstates (recall $\nu$ even):
\begin{equation}
    \begin{aligned}
        \ket{\nu \kl}_p &\approx \frac{\ket{\nu}_B + \ket{\nu-1}_B}{\sqrt{2}} \\
        &\overset{\hat{Z}_\nu}{\longleftrightarrow} \frac{\ket{\nu}_B - \ket{\nu-1}_B}{\sqrt{2}} \approx \ket{-\nu \kl}_p.
    \end{aligned}
\end{equation}
The other gates in Eq.~\eqref{eq:qubitgates} cannot be found in the standard Mach-Zehnder protocol.  The Hadamard gate $\hat{H}_\nu$ transforms between Bloch and momentum bases. It maps the even-parity state $\ket{\nu}_B$, which describes a state with equally split momentum, to the positive momentum state $\ket{\nu \kl}_p$ (approximately) and vice versa. Similarly, it maps between the odd-parity state $\ket{\nu-1}_B$ and $\ket{-\nu \kl}_p$. Therefore, if the atoms are in a split state and propagating apart, the application of $\hat{H}_\nu$ splits each arm $\ket{\pm \nu \kl}_p$ again to then create four arms in total -- two positive momentum and two negative momentum arms.  However, when two arms come together, interfere and an $\hat{H}_\nu$ is applied, they may merge and propagate in only one direction as they were mapped to an (approximate) momentum eigenstate.  

The operations in Eq.~\eqref{eq:qubitgates} are required, for instance, to create that branches of a multi-armed interferometer to measure, for instance, gravity gradients and rotations (c.f.~\cite{trimecheConceptStudyPreliminary2019a, carrazSpaceborneGravityGradiometer2014a}), as well as manipulate the relative phases and populations between arms.  For instance, the $\hat{X}_\nu$ gate performs a $\pi$-rotation around the $x$-axis in the Bloch sphere, which leaves the population in the states $\ket{\pm \nu \kl}_p$ invariant but flips their relative phase. More generally, the design of arbitrary $\hat{R}_{x, \nu}(\phi)$ rotations would allow to impart different relative phases on the interferometer arms in order to, for instance, shift the phase of the interference fringe pattern.

``Partial'' mirror operations can be achieved with $\hat{R}_{z, \nu}(\theta)$ rotation gates as they move population between the momentum states:
\begin{equation}
    \hat{R}_{z, \nu}(\theta) \ket{\pm \nu \kl}_p \approx \cos(\frac{\theta}{2}) \ket{\pm \nu \kl}_p - i \sin(\frac{\theta}{2}) \ket{\mp \nu \kl}_p.
\end{equation}
Clearly $\theta = \pi$ recovers the standard \texttt{MIRROR}$_\nu$, but this also shows the effect of the $\hat{T}_\nu$ gate, which corresponds to $\theta = \pi/4$, resulting in an unbalanced fraction of the interferometer arm ($\sim \sin^2(\pi/8)$) changing direction. The squared operation $\hat{T}_\nu^2 = \sqrt{\hat{Z}_\nu}$ can then be viewed as a ``half'' mirror since $\cos^2(\pi/4) = \sin^2(\pi/4) = 1/2$. Reiterating the statement from above, any qubit gate can be composed from these elementary gates that we include in Eq.~\eqref{eq:qubitgates}, which allows for arbitrary transformations within a conduction band qubit subspace.

In order to optimize the control function producing these gates, we use the infidelity expression in Eq.~\eqref{eq:infid_unitary} with $\hat{W}_\nu = \hat{P}_\nu^\dagger \hat{W}_L \hat{P}_\nu$, where $\hat{W}_L \in SU(2)$ is one of the four operations in Eq.~\eqref{eq:qubitgates} and $\hat{P}_\nu$ is the isometry as introduced in Sec.~\ref{sec:sl}. We obtain the control functions for the lowest conduction band qubit $\hat\Pi_4$ and show the results together with the corresponding state evolutions in Fig.~\ref{fig:xzht}. The bottom row shows the Fourier transformation $\varphi(\omega)$, providing information on the dominant spectral components of the control function.

From Fig.~\ref{fig:xzht}, we note the many distinct state transfer pathways between Bloch states achieved by the control function for each gate.  
The $\hat{X}_4$ gate achieves population inversion in the $\hat{\Pi}_4$ qubit subspace by temporarily transferring to the $\hat\Pi_6$ qubit subspace as well as the lower energy state $\ket{2}_B$. The broad Fourier spectrum reflects the presence of these multiple transitions. The situation is different for the phase gates $\hat{Z}_4$ and $\hat{T}_4$ where specific transitions to the valence band dominate, leading to a dominant Fourier component $\omega \approx 11 \, \omega_R$.  This supports the heuristic usage of this frequency in prior works to design control functions, for instance with reinforcement learning methods \cite{ledesmaDemonstrationProgrammableOptical2024, chihReinforcementlearningbasedMatterwaveInterferometer2021c}. The use of quantum optimal control makes this restrictive design assumption unnecessary, which is one of the advantages as described in Sec. \ref{sec:qoc}. Last, the Hadamard gate $\hat{H}_4$ employs many intermediate Bloch states in order to transform the basis due to a strict phase convention. 

For completeness, and to demonstrate the advantages and generality of the direct collocation method, we want to realize the $\hat{Z}_6$ gate, which mirrors the interferometer arms at a higher momentum $\ket{\pm 6 \kl}_p$. Computationally, this case is more complicated than the conventional mirror in the lowest conduction band qubit ($\hat{Z}_4$) due to the involvement of a larger set of Bloch states.  It is helpful to think about this problem classically to gain intuition for this task. A mirror inverts momentum, thus we aim to perform the transformations $\ket{\pm 2n \kl}_p \rightarrow \ket{\mp 2n \kl}_p$. The operators $\hat{b}_\pm$ can only alter the momentum in steps of $\pm 2 \kl$ ($n \rightarrow n \pm 1$), thus it is reasonable to expect that for larger momenta the complexity of control function achieving the $\hat{Z}_\nu$ (\texttt{MIRROR}) gate increases as a greater momentum difference has to be imparted in gradual steps. Furthermore, it is conceivable that a standard (indirect) optimal control method might struggle at the task as it would only optimize over the control parameters $\varphi_l$ and could not immediately exploit any knowledge about the quantum state trajectory, highlighting the potential advantages of the direct collocation method used in this work.

In particular, classical intuition allows us to form additional constraints for such high-momentum gates to aid in their design.  Classically, there must be a point in time during mirroring where the atoms are stationary. Due to time-reversal symmetry, we expect this to occur at time $T/2$. Since we are designing our gates in the basis of the Bloch states, we are interested in the highest-energy Bloch state that has a significant zero-momentum component. For our choice of lattice depth $V_0 = 10 \, \wr$, this is the state $\ket{2}_B$, which is also the closest in energy to the initial ($\ket{\nu}_B$, $\ket{\nu-1}_B$) and target split states ($\hat{Z}_\nu \ket{\nu}_B$, $\hat{Z}_\nu \ket{\nu-1}_B$) and due to its zero-momentum component allows for transition through the stationary state. This is why one could expect the state $\ket{2}_B$ to be populated at $T/2$.  We incorporate this intuition into an additional constraint on
the time evolution operator $\hat{U}_{l'} = x_{l'}$ at time $t_{l'} = T/2$:
\begin{equation}
   \abs{\bra{2}_B x_{l'} \ket{\nu}_B}^2 \geq 0.1, \quad \abs{\bra{2}_B x_{l'} \ket{\nu-1}_B}^2 \geq 0.1.
\end{equation}
This forces the population of the state $\ket{2}_B$ to be at least $10\%$ at time $T/2$. We choose $10\%$ to inform the optimization that populating this state is required while aiming not to be too restrictive, otherwise the specific value is not of great relevance. It is important to keep in mind that this modification is directly enabled by the chosen optimal control method and would be nontrivial to realize with alternative frameworks (see Refs. \cite{ledesmaDemonstrationProgrammableOptical2024, chihReinforcementlearningbasedMatterwaveInterferometer2021c, shaoApplicationQuantumOptimal2023b, nicotraModelingControlUltracold2023b}).

\begin{figure}
    \centering
    \includegraphics[width=0.9\linewidth]{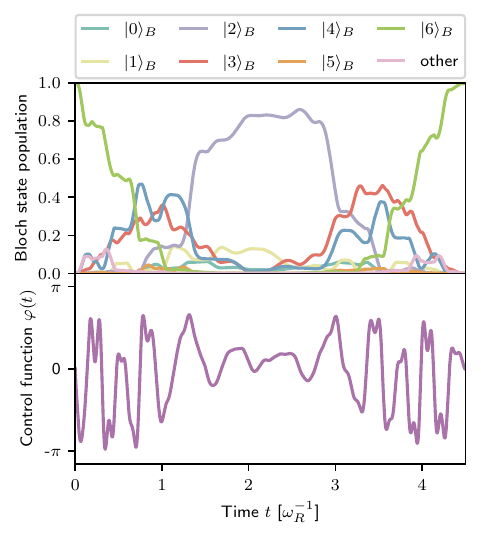}
    \caption{$\hat{Z}_6$ gate acting on qubit $\hat\Pi_6$, $T = 4.50 \, \wr^{-1}$, $\infid = 2.90 \times 10^{-3}$. It is equivalent to a mirror operation for atoms in $\ket{\pm 6 \kl}_p$. The top panel shows the state evolution in the Bloch basis while the lower panel shows the corresponding control function.}
    \label{fig:Z6}
\end{figure}

We now apply this extended problem formulation to design control functions for the $\hat{Z}_6$ gate. Figure~\ref{fig:Z6} shows the state evolution as well as the control function waveform. As expected, a significantly higher gate duration is required, and we note the expected large occupation of the state $\ket{2}_B$ at time $T/2$. The population of this state is much larger than our lower bound of $10\%$, thus verifying our reasoning for implementing the additional constraints.

\subsection{Interqubit gates}
In Sec.~\ref{sec:split}, an example of an operation that maps between qubit subspaces was provided.  This mapping considered only two states with one drawn from each qubit subspace.  One can however imagine a broader set of useful operations $\hat{W} : \hat{\Pi}_\nu \leftrightarrow \hat{\Pi}_{\nu'}$ that map between qubit subspaces.  Within the conduction band, this includes $\texttt{BOOST}_\nu : \hat\Pi_\nu \rightarrow \hat\Pi_{\nu + 2}$ or $\texttt{SLOW}_\nu : \hat\Pi_\nu \rightarrow \hat\Pi_{\nu - 2}$ operations which accelerate or decelerate the atoms, respectively. Furthermore, a mapping between conduction and valence band qubit subspaces can serve to \texttt{HOLD} or \texttt{RELEASE} atoms, for instance as steps in a lattice hold procedure designed to increase the space-time area enclosed within a spatially confined system (c.f.~\cite{xuProbingGravityHolding2019a}). In the present work, we discuss results for these operations connecting the qubit subspaces $\hat\Pi_1$ and $\hat\Pi_4$.


In order to optimize a subspace transfer we propagate the initial qubit subspace $\ket{\psi_0(t=0)} = \ket{\nu}_B$ and $|\psi_1(t=0)\rangle = \ket{\nu - 1}_B$ and project the final states into the target qubit subspace $\hat\Pi_{\nu'}$. The infidelity 
\begin{equation}
    \label{eq:infid_transfer}
    \infid = 1 - \frac{1}{2} \sum_{k=0}^1 \bra{\psi_k(T)} \hat{\Pi}_{\nu'} \ket{\psi_k(T)}
\end{equation}
describes the amount of population that is not transferred into the desired subspace, thus $\infid = 0$ corresponds to a perfect transfer. For this operation it is not relevant where exactly in the target Bloch sphere the states in the initial qubit subspace are mapped to; when the transfer operation is optimized and fully characterized, additional corrections inside the target qubit subspace can be performed via single-qubit $SU(2)$ gates, which were discussed in Sec.~\ref{sec:qubitgate}.

\begin{figure}
    \centering
    \includegraphics[width=0.9\linewidth]{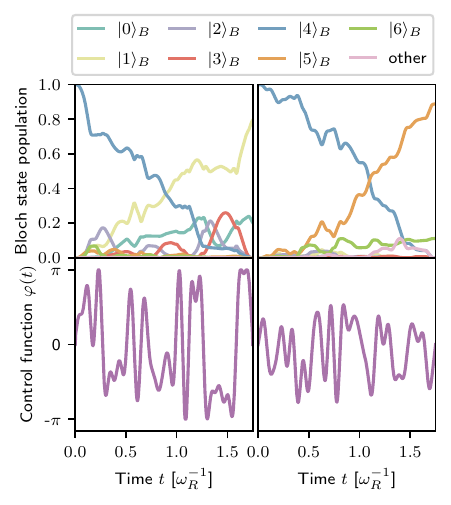}
    \caption{Control functions for the \texttt{HOLD} operation (left, $\infid = 1.37 \times 10^{-3}$) and the \texttt{BOOST}$_4$ operation (right, $\infid=9.13 \times 10^{-5}$). For both operations the gate time is $T=1.75 \, \omega_r^{-1}$.}
    \label{fig:holdboost}
\end{figure}

In Fig.~\ref{fig:holdboost} we show the results for the $\texttt{BOOST}_4$ and \texttt{HOLD} operations. As expected, when evolving a basis state of the initial manifold we do not arrive at a basis state of the target manifold but instead in a superposition. This is due to the particular choice of infidelity function in Eq.~\eqref{eq:infid_transfer} which only fixes the target manifold but not a particular state. It allows for more freedom for the optimization compared to optimizing for a fixed unitary gate acting on both subspaces. We note that the related inverse actions $\texttt{SLOW}_6$ and \texttt{RELEASE} can be obtained via time-reversal symmetry of the corresponding control functions.

\section{Outlook \& Conclusion}\label{sec:conclusion}

In this work we have demonstrated how the direct collocation quantum optimal control method can be used to design control functions for operations relevant to inertial sensing on various qubit subspaces of an atom confined in a one-dimensional optical lattice.  We presented a framework that breaks the complex control space of the optical lattice interferometer architecture into fundamental computational units that can be acted on by quantum gates in analogy to quantum computing. We identified two-level subspaces which are embedded in the infinite-dimensional optical lattice system and can be thought of as logical qubit subspaces that describe a specific momentum configuration of atoms in the lattice. This work demonstrated the design of control functions that implement physically motivated transformations on and between these logical qubit subspaces.

\begin{figure}
    \centering
    \includegraphics[width=\linewidth]{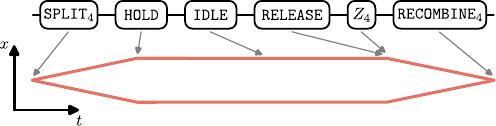}
    \caption{Space-time representation of a simple circuit for a lattice-hold interferometer sequence written using the gates proposed in this work.}
    \label{fig:example_circ}
\end{figure}

The larger aim of this work is to introduce ``programmability'' into this platform by gathering a catalog of useful operations, which allow abstraction and consequently sensing-protocol design which rises above the level of the control function dynamics themselves. This has a direct analogy to the kinds of abstractions which occur in the field of digital quantum computing where algorithms are expressed in terms of quantum circuits.  Using such primitives, various sensing protocols for the optical lattice interferometer can be expressed in the circuit language.  Figure~\ref{fig:example_circ} shows an example of a circuit of gates that realize an interferometry protocol where the atom cloud is split, propagated, held, mirrored and eventually recombined, allowing for the enclosure of a large space-time area (and thus high sensitivity) without increasing the spatial size of the lattice.  More complex protocols that involve interqubit operations and qubit subspace operations can be developed without having to return to the (non-intuitive) level of the control function and Bloch state transfer dynamics. 

Much as quantum computing, this transition from pulse to circuit level design has obvious practical advantages also for quantum sensing.  In the sensing of spin-rotations this transition was achieved already through the direct utilization of gatesets already available on digital quantum computers as programmable quantum sensors.  For inertial sensors, one can easily imagine a catalog of operations (those found in Sec.~\ref{sec:opt} constituting only a small subset) that can be used to assemble sensing protocols in one, two, and three dimensions in order to sense vector components of acceleration, rotation, gravity, and force gradients (c.f.~Ref.~\cite{ledesmaVectorAtomAccelerometry2024a,wuGravitySurveysUsing2019a}). This also includes calibrating protocols which are sensitive to internal settings of a device that are subject to drift as well as the changing inertial and electromagnetic environment of a device (c.f.~\cite{alamRobustQuantumSensing2024a}). Just like in quantum computing, one could imagine utilizing the catalog of operations available on an inertial sensor in order to characterize and verify performance of a device \cite{hashimPracticalIntroductionBenchmarking2024a}.  This could include  the equivalent of mid-circuit measurements via (for instance) non-destructive measurements or the transfer (outcoupling) of atomic populations into untrapped states \cite{blochAtomLaserCw1999a, serafiniVortexReconnectionsRebounds2017a}.  Here the prospect of correcting errors in quantum sensors has been the subject of recent interest \cite{zhouAchievingHeisenbergLimit2018a, kesslerQuantumErrorCorrection2014a}, highlighting the need for protocols to generate quantum resources (such as entanglement and squeezing) that are useful for inertial sensing (c.f.~\cite{szigetiImprovingColdatomSensors2021a, greveEntanglementenhancedMatterwaveInterferometry2022a, wilsonEntangledMatterWaves2024a, cassensEntanglementenhancedAtomicGravimeter2024a, brennenQuantumLogicGates1999a, kendellErrorCorrectingStates2023a}).

\section{Acknowledgements}

We acknowledge fruitful discussions on optical lattice interferometry with Murray Holland, John Wilson, Shah Saad Alam, Noah Fitch, Dana Anderson, and Zachary Pagel.  
This work is funded in part by STAQ under award NSF Phy-232580 and in part by the US Department of Energy Office of Advanced Scientific Computing Research, Accelerated Research for Quantum Computing Program. This material is supported by the U.S. Department of Energy, Office of Science, Office of Advanced Scientific Computing Research under Award Number DE-SC0021526, and under Contract No.~DE-AC02-06CH11357 of the Exploratory Research for Extreme-Scale Science program.

{\it Note added} -- In the preparation of this work for submission, we became aware of related work from {\it LeDesma et al.} \cite{ledesmaUniversalGateSet2024a}, studying gate sets as well as their implementation on an optical lattice interferometer.

\bibliography{references2}

\end{document}